\documentstyle[aps,twocolumn,floats,graphicx]{revtex}

\begin{document}
\title{{\bf Supercriticality and Transmission Resonances in the Dirac Equation}}
\author{{\bf Norman Dombey}
$^{*a}$, {\bf Piers Kennedy}$^{+a}${\bf \ and A.
Calogeracos }$^{\dagger b}$}
\address{$^a$Centre for Theoretical Physics, University of Sussex, 
Brighton BN1 9QJ,~UK}
\address{$^b$PO Box 61147, Maroussi 151 22,~Greece\\
email:$^{*}$normand@sussex.ac.uk; $^{+}$kapv4@pcss.maps.susx.ac.uk;
$^{\dagger}$acal@hol.gr}
\address{\flushleft\rm 
It is shown that a Dirac particle of mass $m$ and arbitrarily small momentum will tunnel without
reflection through a potential barrier $V=U_c(x)$ of finite range provided that the
potential well $V=-U_c(x)$ supports a bound state of energy $E=-m.$ This is called
a supercritical potential well.}
\date{SUSX-TH/00-011}
\maketitle

It is now over 70 years since the Dirac equation was written down. Yet new
results have been discussed in recent years, even in the relatively simple
cases of one \cite{one} and two spatial dimensions \cite{two} as well as in
three dimensions \cite{three}-\cite{six}. In this note, we generalise a well
known theorem of scattering off a one-dimensional potential well  in the
Schr\"{o}dinger equation to the Dirac equation. This is not difficult, but the
theorem has an unexpected twist. Since the Dirac equation covers
anti-particle scattering as well as particle scattering, the generalisation
gives two distinct results. One of these results implies a remarkable
property of tunnelling through a potential barrier in the Dirac equation
which is related to the result on barrier penetration found by Klein \cite
{klein} and now called the Klein Paradox.

We begin by considering the scattering off a class of one-dimensional
potential wells $V(x)$ where $V(x)=0$ for $\left| x\right| \geq a$ and $%
V(x)=-U(x)$ $\leq 0$ for $\left| x\right| <a$ where the piecewise continuous
function $U(x)\geq 0$ . The potential is also taken to be even  so that $%
V(-x)=V(x).$ We first seek to generalise to the Dirac equation the
non-relativistic result that  the reflection coefficient $R(k)$ for
scattering off the potential well $V(x)=-U_0(x)$ which supports a zero
energy resonance satisfies $R(0)=0$, where $k$ is the momentum of the
particle. This theorem was known to Schiff \cite{schiff} and Bohm \cite{bohm}
but a proof was published only  relatively recently by Senn \cite{senn} and
Sassoli de Bianchi \cite{others}. The situation where $R(k)=0$ and the
transmission coefficient $T(k)=1$ is called a transmission resonance \cite
{bohm}. In non-relativistic systems a zero energy resonance (or half-bound
state) \cite{newton} is the non-trivial limit where a bound state just
emerges from the continuum, for example when a square well potential is just
strong enough to support a second bound state.

Following an earlier paper\cite{cdi} we take the gamma matrices $\gamma _x$
and $\gamma _0$ to be the Pauli matrices $\sigma _x$ and $\sigma _z$
respectively. Then the Dirac equation for scattering of a particle of energy 
$E$ and momentum $k$ by the potential $V(x)$ can be written as the coupled
equations

\begin{equation}
\frac{\partial f}{\partial x}+(E-V(x)+m)g=0  \eqnum{1a}
\end{equation}

\begin{equation}
\frac{\partial g}{\partial x}-(E-V(x)-m)f=0  \eqnum{1b}
\end{equation}

\noindent 
where the Dirac spinor $\psi =\left( 
\begin{array}{c}
f \\ 
g
\end{array}
\right) .$

Eqs. (1) have simple solutions as $x\rightarrow \pm \infty $ where $V=0.$ In
particular, the analogue of a zero energy resonance in the Schr\"{o}dinger
equation is a zero momentum resonance in the Dirac equation \cite{two} where
a particle of zero momentum has $E=m$ or an anti-particle has $E=-m$.%
\footnote{%
The anti-particle is described by the hole wave function corresponding to
the absence of the state with $E=-m$}. It is easy to see that the solution
of Eq. (1) for $E=m$ and $V=0$ appropriately normalised is $\psi =\left( 
\begin{array}{c}
2m \\ 
0
\end{array}
\right) $ while the solution for $E=-m$ and $V=0$ is $\left( 
\begin{array}{c}
0 \\ 
2m
\end{array}
\right) $. As in Ref. \cite{cdi} we can now write down the solutions of Eq.
(1) for a particle of momentum $k$ as $x\rightarrow \pm \infty $ to obtain $%
\psi =$ $\left( 
\begin{array}{c}
E+m \\ 
-ik
\end{array}
\right) e^{ikx}$ while an anti-particle of momentum $k$ will have $\psi =$ $%
\left( 
\begin{array}{c}
ik \\ 
m-E
\end{array}
\right) e^{-ikx}$

We now set up the usual formalism for particle scattering by the potential $%
V(x)$ in the Dirac equation. We take the particle as incident from the left,
so the amplitude for reflection $r(k)$ is defined through the spinor $\psi
(x)$ as $x\rightarrow -\infty $

\begin{equation}
\psi (x)=\left( 
\begin{array}{c}
E+m \\ 
-ik
\end{array}
\right) e^{ikx}+r(k)\left( 
\begin{array}{c}
E+m \\ 
ik
\end{array}
\right) e^{-ikx}  \eqnum{2}
\end{equation}

\noindent while as $x\rightarrow \infty $

\begin{equation}
\psi (x)=t(k)\left( 
\begin{array}{c}
E+m \\ 
-ik
\end{array}
\right) e^{ikx}  \eqnum{3}
\end{equation}

In the Dirac equation \cite{thal} as for the Schr\"{o}dinger equation with
symmetric potentials \cite{others}, unitarity implies that

\begin{equation}
\left| r\right| ^2+\left| t\right| ^2=1;\qquad Im(r^{*}t)=0  \eqnum{4}
\end{equation}

\noindent so $R+T=1$ where the reflection and transmission coefficients are
given by $R=\left| r\right| ^2,T=\left| t\right| ^2.$

Since the potentials we consider are even, parity is conserved. In our
two-component approach, the transformation of a wave function 
under $x\rightarrow -x$
is given \cite{cdi} by

\begin{equation}
\psi^{\prime} (x,t)=\sigma_z \psi (-x,t)  \eqnum{5}
\end{equation}

\noindent  It follows that an even wave function $\psi $ has an even top
component $f$ and an odd bottom component $g$. Similarly for an
odd wave function $\psi ,$ $f$ will be odd and $g$ will be even.

We first consider an even bound state in the potential $V(x).$ As $%
x\rightarrow \pm \infty ,$ its unnormalised wave function will be of the form

\begin{equation}
\psi (x)=\left( 
\begin{array}{c}
m+E \\ 
-\kappa
\end{array}
\right) e^{\kappa x} \quad x \rightarrow -\infty  \eqnum{6a}
\end{equation}

\begin{equation}
\psi (x)=\left( 
\begin{array}{c}
m+E \\ 
\kappa
\end{array}
\right) e^{-\kappa x} \quad x \rightarrow \infty  \eqnum{6b}
\end{equation}

\noindent where $E^2=m^2-\kappa ^2$. We require the potential well $%
V=-U_0(x) $ to just bind this bound state with arbitrarily small $\kappa $.
If this is the case, then the limit $\kappa \rightarrow 0$ exists, and $\psi
(x)$ becomes a continuum wave function since it is no longer square
integrable.

We can now compare Eqs. (6) with Eqs (2) and (3) in the limit $k\rightarrow
0,\kappa \rightarrow 0.$ We obtain $2m(1+r(0))=2mt(0)$ or

\begin{equation}
1+r(0)=t(0)\neq 0  \eqnum{7}
\end{equation}

\noindent We have written $t(0)\neq 0$ since otherwise $\psi (x)$ would
vanish in the limit $k\rightarrow 0,\kappa \rightarrow 0$ and we would not be
considering a zero momentum resonance. \footnote{%
In Ref \cite{piers} we adopt a more general approach to obtain the results
of this letter thereby avoiding the use of
$t(0)\neq 0:$}

The theorem now follows easily just as it does in the Schr\"{o}dinger case \cite
{others}. Combining Eq (7) with the unitarity condition $Im(r^{*}t)=0$ of Eq
(4), we get $Im(r^{*})=0$ so that $r(0)$ and $t(0)$ are real. From $\left|
r\right| ^2+\left| t\right| ^2=1$ we obtain

\begin{equation}
2r^2+2r+1=1  \eqnum{8}
\end{equation}

\noindent so that $r(0)=0$ or $r(0)=-1.$ Since $t(0)\neq 0$ we obtain the
result $r(0)=0$ and so in terms of the reflection and transmission
coefficients

\begin{equation}
R(0)=0\qquad T(0)=1  \eqnum{9}
\end{equation}

If instead we had considered an odd bound state, an additional minus sign
must be introduced into either Eq. (6a) or Eq. (6b). Eq. (7) must be
modified to $1+r(0)=-t(0)$ and the subsequent analysis and conclusions
remain valid. Hence just as in the Schr\"{o}dinger equation, the scattering of a
particle in the Dirac equation off a potential well $V=-U_0(x)$ which \lq\lq
binds'' a zero momentum resonance corresponds to a transmission resonance
with zero reflection.

We now increase the strength of the potential well from $U_0(x)$ to $U_c(x)$
so that $V=-U_c(x)$ supports a bound state of energy $E=-m.$ This is called
a supercritical potential and is associated with spontaneous positron
production \cite{grein}, \cite{zeld}. We can redo the analysis exactly as
before by defining amplitudes $r_{-},t_{-}$ for the reflection and
transmission of an anti-particle of momentum $k$ incident from the left on a
potential well $V(x)$: so in place of Eq. (2) we have as $x\rightarrow
-\infty $

\begin{equation}
\psi (x)=\left( 
\begin{array}{c}
ik \\ 
m-E
\end{array}
\right) e^{-ikx}+r_{-}(k)\left( 
\begin{array}{c}
-ik \\ 
m-E
\end{array}
\right) e^{ikx}  \eqnum{10}
\end{equation}

\noindent while as $x\rightarrow \infty ,$ we have

\begin{equation}
\psi (x)=t_{-}(k)\left( 
\begin{array}{c}
ik \\ 
m-E
\end{array}
\right) e^{-ikx}  \eqnum{11}
\end{equation}

\noindent As before unitarity gives

\begin{equation}
\left| r_{-}\right| ^2+\left| t_{-}\right| ^2=1;\qquad Im(r_{-}^{*}t_{-})=0 
\eqnum{12}
\end{equation}

\noindent and the near-supercritical even bound state for $x\rightarrow \pm
\infty $ is now

\begin{equation}
\psi (x)=\,\left( 
\begin{array}{c}
-\kappa \\ 
m-E
\end{array}
\right) e^{\kappa x}\quad x\rightarrow -\infty  \eqnum{13a}
\end{equation}

\begin{equation}
\psi (x)=-\left( 
\begin{array}{c}
\kappa \\ 
m-E
\end{array}
\right) e^{-\kappa x}\quad x\rightarrow \infty  \eqnum{13b}
\end{equation}

\noindent Note again that for the odd bound state we must drop the minus
sign in Eq. (13b).

Repeating the analysis of Eqs (7-9) we find in the limit $k\rightarrow
0,\kappa \rightarrow 0$  when the antiparticle is incident on the potential
well $V=-U_c(x)$ with arbitrarily small momentum that

\begin{equation}
R_{-}(0)=0\qquad T_{-}(0)=1  \eqnum{14}
\end{equation}

\noindent where $R_{-}=\left| r_{-}\right| ^2,T_{-}=\left| t_{-}\right| ^2$

So we see that in the Dirac equation there are two analogues of the
Schr\"{o}dinger result: one for zero momentum particles incident on a potential
well which supports a zero momentum resonance and one for zero momentum
particles incident on a supercritical potential well.

We now can obtain our main result. The Dirac equation (1) is invariant under
charge conjugation: that is to say under the transformation

\begin{equation}
E\rightarrow -E\quad V\rightarrow -V\quad f\rightarrow g\quad g\rightarrow f
\eqnum{15}
\end{equation}

From Eq (14) we know that an antiparticle of energy $E=-\sqrt{m^2+k^2\text{ }%
}$incident on the supercritical potential well $V_c(x)=-U_c(x)$ will satisfy 
$T_{-}(0)=1$, that is to say at arbitrarily small momentum it will have a
vanishingly small reflection coefficient. Eq. (14) then shows that if we
replace the antiparticle of energy $E=-\sqrt{m^2+k^2\text{ }}$ incident on
the supercritical potential well by a particle of energy $E=\sqrt{m^2+k^2}$
incident on the corresponding  potential barrier $V(x)=+U_c(x)$ the particle
will still have a transmission resonance at zero momentum, even though now
the potential well has been replaced by a potential barrier.We thus obtain
the theorem that where an even potential well of finite range is strong
enough to contain a supercritical state, then a particle of arbitrarily
small momentum will be able to tunnel right through the potential barrier
created by inverting the well without reflection.This result was noticed a
few years ago for the particular case of square barriers \cite{cdi}, and one
of us (PK) has shown numerically that it was also true for Gaussian and
Saxon-Woods potential barriers \cite{piers2}. In the Appendix we show the
behaviour of the upper and lower components of the wave function for
particle scattering at zero momentum by a square and Gaussian barrier when
the corresponding potential wells are supercritical.

{\bf Conclusions}

In his original work Klein \cite{klein} discovered that a Dirac particle
could tunnel through an arbitrarily high potential. The generic phenomenon
whereby fermions can tunnel through barriers without exponential suppression
we have called \lq\lq Klein Tunnelling'' \cite{cd}. The result of this letter
shows that Klein tunnelling is a general feature of the Dirac equation: any
potential well strong enough to support a supercritical state when inverted
becomes a potential barrier which a fermion of arbitrarily low momentum can
tunnel through without reflection. We do not claim here that any
transmission resonance at zero momentum must correspond to supercriticality,
only that supercriticality leads to a transmission resonance through a
potential barrier at zero momentum. In another paper \cite{piers} we shall 
consider the question of the conditions on a potential for it to
possess a zero momentum transmission resonance more generally. In three
dimensions Hall and one 
of us (ND) \cite{hall} have recently demonstrated  that 
maximal Klein tunnelling is also associated with supercriticality. 

The potential step that Klein considered has pathological properties \cite
{cd}. Nevertheless our result confirms that according to the Dirac equation
a particle of low momentum can tunnel through an arbitrarily high smooth
potential of finite range. The reason is straightforward: hole states can
propagate under the potential barrier. In terms of the particle kinetic
energy T$\,$under the barrier T$\,=E-V-m=-m-\sqrt{m^2+q^2}$ where $q$ is the
momentum of the hole so if T$\,\leq -2m$, hole states can propagate without
exponential suppression. T$\,\leq -2m$ thus corresponds to penetrating under
the barrier to distances $\left| x\right| <\left| x_K\right| $ where $%
V(x_K)=E+m\geq 2m$ \cite{hall}.

{\bf Appendix}

We illustrate the result above for the special cases of (i) a square barrier
and (ii) a Gaussian barrier. First consider the square well potential $%
V=-U(x)$ where $U(x)=U$ for $\left| x\right| <a$ and $U(x)=0$ for $\left|
x\right| >a.$ Then an unnormalised even wave function inside the well \cite
{cdi} has the form

\begin{equation}
\psi (x)=\left( 
\begin{array}{c}
(E+U+m)\cos px \\ 
p\sin px
\end{array}
\right) \quad |x|\leq a  \eqnum{16}
\end{equation}

\noindent where the internal momentum $p$ is given by $(E+U)^2=m^2+p^2$. For
supercriticality where $E=-m$ we require the phase condition $pa=N\pi /2$
where $N$ is an integer \cite{cdi}. The first supercritical state is thus
given by $p=p_c=\pi /2a$ and correspondingly the critical potential is $%
V(x)=-U_c\quad |x|\leq a,$ where $U_c=m+\sqrt{m^2+\pi ^2/4a^2}$ At
supercriticality the wave function is thus given by

\begin{equation}
\psi (x)=\left( 
\begin{array}{c}
0 \\ 
2m
\end{array}
\right), \quad x<-a  \eqnum{17a}
\end{equation}

\begin{equation}
\psi (x)=-2m\left( 
\begin{array}{c}
b\cos (\pi x/2a) \\ 
sin(\pi x/2a)
\end{array}
\right), \quad |x|\leq a  \eqnum{17b}
\end{equation}

\begin{equation}
\psi (x)=-\left( 
\begin{array}{c}
0 \\ 
2m
\end{array}
\right), \quad x>a  \eqnum{17c}
\end{equation}

\noindent where $b=2aU_c/\pi .$

Now consider a particle of arbitrarily small momentum incident on the square
barrier $V(x)=U_c\quad |x|<a;\,V(x)=0\quad |x|>a.$ Eq (15) shows that the
wave function is obtained by interchanging the top and bottom components of
Eq (17) thereby giving the transmission resonance

\begin{equation}
\psi (x)=\left( 
\begin{array}{c}
2m \\ 
0
\end{array}
\right), \quad x<-a  \eqnum{18a}
\end{equation}

\begin{equation}
\psi (x)=-2m\left( 
\begin{array}{c}
\sin (\pi x/2a) \\ 
b\cos (\pi x/2a)
\end{array}
\right), \quad |x|\leq a  \eqnum{18b}
\end{equation}

\begin{equation}
\psi (x)=-\left( 
\begin{array}{c}
2m \\ 
0
\end{array}
\right), \quad x>a  \eqnum{18c}
\end{equation}

\noindent and the components of the wave function are shown in Fig.1. 
\begin{figure}[tbph]
\par
\begin{center}
\leavevmode
\includegraphics[width=0.7\linewidth]{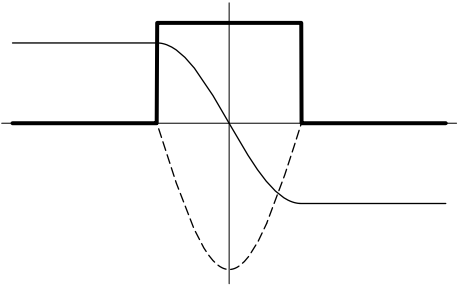} \medskip
\end{center}
\caption{The zero momentum wave function for the square barrier $V=U_c(x)$,
depicted by the heavy line. The solid line is the upper component and the
dashed line is the lower component.}
\end{figure}

One of us (PK) has also solved the Dirac equation numerically for a Gaussian
potential well and barrier where $U(x)=U\exp (-x^2/a^2)$ \cite{piers2}. In
Fig. 2 we show the components of the wave function for a particle of
arbitrarily small momentum incident on a supercritical Gaussian barrier
where $U=U_c=3.26m$ for $ma=1$ (cf. $U_c=m+\sqrt{m^2+\pi ^2/4a^2}=2.86m$ for 
$ma=1$ for a square barrier). Again there is a transmission resonance
demonstrating complete penetration of the barrier.

\begin{figure}[htbp]
\par
\begin{center}
\leavevmode
\includegraphics[width=0.7\linewidth]{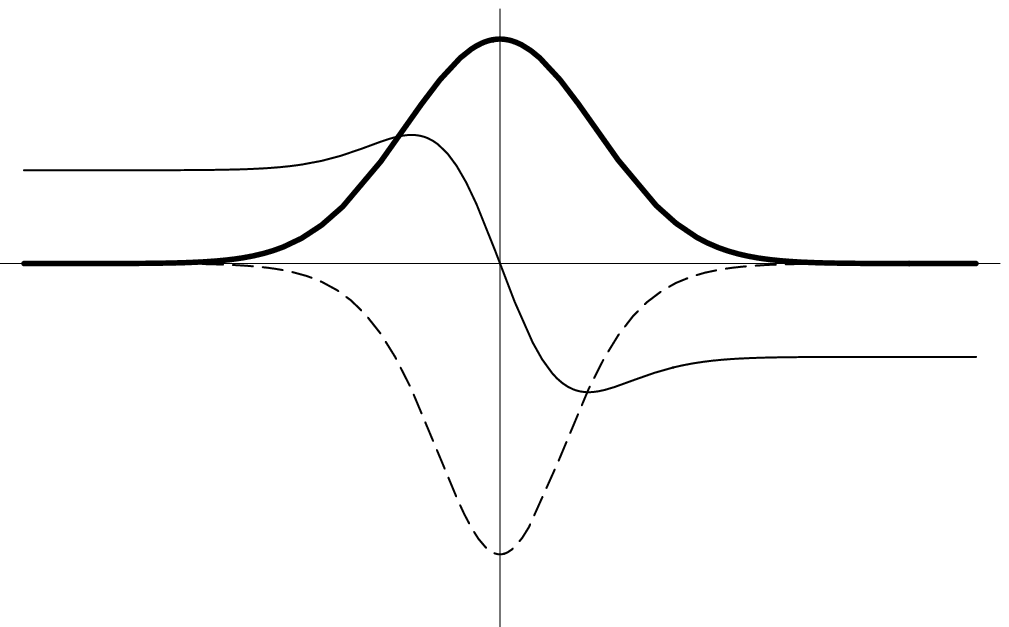} \medskip
\end{center}
\caption{The zero momentum wave function for the Gaussian barrier $V=U_c\exp
(-m^2x^2)$, depicted by the heavy line. The solid line is the upper
component and the dashed line is the lower component}
\end{figure}

While the wave functions in this case have a similar form to those for the
square barrier, note the two turning points which occur in the top component
of the Gaussian wave function. These correspond to the points $\pm x_K$
where $V(x_K)=E+m=2m$ at zero momentum. Hole states can propagate under
the potential without exponential suppression from $-x_K$ to $+x_K$ thus
demonstrating Klein tunnelling. Note also that the condition 
for hole states to propagate under a potential of finite range  
is $V>2m$ which will in general not
be sufficient for supercriticality (we have seen for a square barrier
of range $a$ with $%
ma=1 $ that $V_c=2.86m$). So Klein tunnelling should exist 
even for subcritical potentials as was pointed out by Jensen et al
\cite{jens}.


\begin{references}
\bibitem{one}  Y Nogami and F M Toyama, Phys Rev ${\bf A57\,}93(1998)$

\bibitem{two}  S H Dong, X W Hou and Z Q Ma, Phys Rev ${\bf A58\,}2160(1998)$

\bibitem{three}  N Poliatzky, Phys Rev Lett ${\bf 70\,}2507(1993);{\bf 76\,}%
3655(1996)$

\bibitem{four}  R G Newton, Helv Phys Acta ${\bf 67\,}20(1994)$

\bibitem{five}  Z Q Ma, Phys Rev Lett ${\bf 76\,}3654(1996)$

\bibitem{six}  R L Hall, Phys Rev Lett ${\bf 83\,}468(1999)$

\bibitem{klein}  O.Klein, Z.Phys. ${\bf 53\,}157\,(1929)$

\bibitem{schiff}  L Schiff, {\it Quantum Mechanics }, (McGraw Hill, New
York) 1949, p. 113

\bibitem{bohm}  D Bohm, {\it Quantum Mechanics}, (Prentice Hall, Englewood
Cliffs, NJ) 1951, p. 286

\bibitem{senn} P Senn, Am. J. Phys. ${\bf 56},916,(1988)$.

\bibitem{others}  M Sassoli de Bianchi, J. Math. Phys. ${\bf 35},2719,(1994)$%
.

\bibitem{newton}  R G Newton, {\it Scattering Theory of Waves and Particles}%
, (Springer-Verlag, Berlin 1982), p.280

\bibitem{cdi}  A Calogeracos, N Dombey and K Imagawa, Yadernaya Fiz. ${\bf %
159\,}1331(1996)$; Phys. At. Nuc.${\bf 159}\,1275(1996)$

\bibitem{thal}  B Thaller, {\it The Dirac Equation}, (Springer-Verlag,
Berlin) 1992, p. 121

\bibitem{piers}  N Dombey and P Kennedy, Transmission Resonances and
Supercriticality in the One Dimensional Dirac Equation, Sussex Preprint
SUSX-TH/007 (to be published)

\bibitem{grein}  W Pieper and W Greiner, Z. Phys.${\bf 218,}327\,(1969)$

\bibitem{zeld}  Ya B\ Zeldovich and S S Gershtein, Zh. Eksp. Teor. Fiz. $%
{\bf 57}\,654(1969)$

\bibitem{piers2}  P Kennedy, A Numerical Investigation of Potential Wells
and Barriers in the One-Dimensional Dirac Equation, M. Sc. dissertation,
University of Sussex (1999)

\bibitem{cd}  A Calogeracos and N Dombey, Int J Mod Phys ${\bf A14\,}%
631(1999)$; Phys. Rep.${\bf 315\,}\,41(1999)$

\bibitem{hall} N Dombey and R L Hall, Phys. Lett. ${\bf 474\,}\,1(2000)$

\bibitem{jens}   H G Dosch, J H D Jensen and V F Muller, Phys. Norvegica $%
{\bf 5},151(1971)$
\end{references}
\end{document}